\documentclass[sigconf,authorversion,screen]{acmart}

\AtBeginDocument{%
  \providecommand\BibTeX{{%
    \normalfont B\kern-0.5em{\scshape i\kern-0.25em b}\kern-0.8em\TeX}}}

\copyrightyear{2020} 
\acmYear{2020} 
\setcopyright{acmcopyright}\acmConference[ICMI '20]{Proceedings of the 2020 International Conference on Multimodal Interaction}{October 25--29, 2020}{Virtual event, Netherlands}
\acmBooktitle{Proceedings of the 2020 International Conference on Multimodal Interaction (ICMI '20), October 25--29, 2020, Virtual event, Netherlands}
\acmPrice{15.00}
\acmDOI{10.1145/3382507.3418885}
\acmISBN{978-1-4503-7581-8/20/10}

\usepackage{graphics} 
\usepackage{epsfig} 
\usepackage{amsmath} 
\usepackage{amssymb}  

\usepackage{multirow}
\usepackage{subfigure}
\usepackage{multirow}
\usepackage{tabu}
\usepackage{mathtools}

\newcommand{\pr}{p}

\begin{document}
\fancyhead{}
\title{Toward Adaptive Trust Calibration for Level 2 Driving Automation}


\author{Kumar Akash}
\email{kakash@purdue.edu}
\orcid{0000-0003-2807-0943}
\affiliation{%
  \institution{Purdue University}
  \city{West Lafayette}
  \state{IN}
  \country{USA}
  \postcode{47907}
}

\author{Neera Jain}
\email{neerajain@purdue.edu}
\affiliation{%
  \institution{Purdue University}
  \city{West Lafayette}
  \state{IN}
  \country{USA}
  \postcode{47907}
}

\author{Teruhisa Misu}
\email{tmisu@honda-ri.com}
\affiliation{%
 \institution{Honda Research Institute USA, Inc.}
 \streetaddress{Rono-Hills}
 \city{San Jose}
 \state{CA}
 \country{USA}
 \postcode{95134}}


\begin{abstract}
Properly calibrated human trust is essential for successful interaction between humans and automation. However, while human trust calibration can be improved by increased automation transparency, too much transparency can overwhelm human workload. To address this tradeoff, we present a probabilistic framework using a partially observable Markov decision process (POMDP) for modeling the coupled trust-workload dynamics of human behavior in an action-automation context. We specifically consider hands-off Level 2 driving automation in a city environment involving multiple intersections where the human chooses whether or not to rely on the automation. We consider automation reliability, automation transparency, and scene complexity, along with human reliance and eye-gaze behavior, to model the dynamics of human trust and workload. We demonstrate that our model framework can appropriately vary automation transparency based on real-time human trust and workload belief estimates to achieve trust calibration. 

\end{abstract}



\begin{teaserfigure}
\centering
\includegraphics[width=\textwidth]{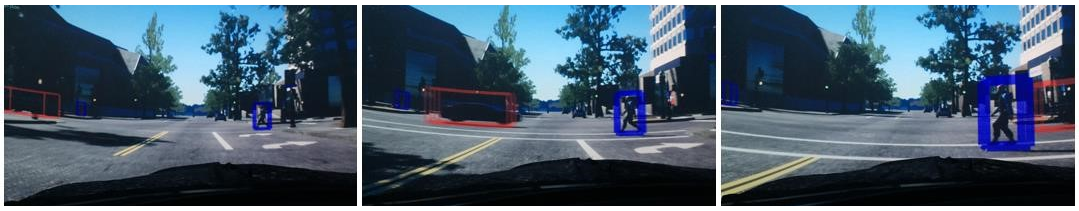}
\caption{Simulation environment of our user study with augmented reality (AR)-based information presentation. The goal of this research is to calibrate a driver's trust in driving automation through AR-based information while avoiding increased driver workload.}
\Description{Simulation environment of our user study with augmented reality (AR)-based information presentation. The goal of this research is to calibrate a driver's trust in driving automation through AR-based information while avoiding increased driver workload.}
\label{fig_ar_example}
\end{teaserfigure}

\maketitle

\section{Introduction} \label{sec_intro}

Humans are increasingly becoming dependent on automation. In the driving domain, advanced driver-assistance systems (ADAS) like adaptive cruise control, lane assist, and collision avoidance have been developed and deployed extensively to vehicles driving on the roads today. Despite significant advancements in these technologies, though, human supervision and intervention are still required. Researchers have shown that human trust plays a critical role in interactions between humans and automated systems. For example, low levels of trust can lead to disuse of automation \cite{choi2015investigating,ghazizadeh2012extending}, whereas excessively relying on the automation capabilities under unsafe conditions, or situations outside of the scope of automation design, can lead to overtrust and consequently, accidents~\cite{muir1987trust,lee2004trust}. Therefore, the goal of the study and resulting framework presented in this paper is to align human trust with the automation's reliability, rather than to maximize human trust.
 
Researchers have proposed to develop paradigms that anticipate human interaction behaviors---such as trust in automation---and influence humans to make optimal choices about automation use \cite{drnec2016paradigm, metcalfe2017building, akash2018classification, hu2018computational}. Pre-requisites for such an approach involve the capability to quantitatively predict human behavior and an algorithm for determining the optimal intervention to influence human behavior. 
Chen et al.~\cite{chen2018planning} has modeled human trust dynamics on a table-clearing task, and adjusted manipulation robot's control behavior considering human trust status. While many studies have optimized the system physical behavior, it is not necessarily an easy approach when the system deals with safety critical tasks. The system must guarantee safety while adapting the control behavior for improved performance, which makes the optimization complicated. Several studies have shown that optimizing the amount of information the system provides (disclosure of the system's internal states) can also achieve a trust calibration without changing system's physical behavior. 
Studies such as \cite{akash2019improving-1,akash2019improving,akash2020human,okamura2018adaptive,tintarev2007survey,felfernig2006empirical,sinha2002role} have shown that the optimization of automation \emph{transparency}\footnote{Following those previous studies we define transparency as ``the descriptive quality of an interface pertaining to its abilities to afford an operator's comprehension about an intelligent agent's intent, performance, future plans, and reasoning process''~\cite{chen2014situation}.} can also contribute to the better collaboration performance.
However, while more information is typically communicated to the human to achieve greater transparency, it often results in increased cognitive workload \cite{tintarev2007survey, helldin2014transparency} and can distract the human from the most critical information \cite{ananny2018seeing} as well as sacrifice one of the primary benefits of the automation, i.e., reduction of human workload. Therefore, a tradeoff between increased trust and increased workload exists when considering increased transparency \cite{alonso2018system}. 

Existing decision-making frameworks do not explicitly model the human workload dynamics required to address this tradeoff \cite{metcalfe2017building,chen2018planning}, in particular for driving contexts. 
This is probably because most above-mentioned studies dealt with decision-aid contexts \cite{akash2019improving-1,akash2019improving} in which the automation only makes a recommendation, and the final action is taken by the human, thus more transparency was usually the better policy. However, several real-life contexts can be classified as `action-automation' where the automation takes the action, unless intervened upon by the human. Examples of the action-automation include automated power plant operation, aircraft autopilot, and self-driving cars. Unlike decision automation, where the humans' interaction with the automation is characterized by their surveillance and compliance whenever the automation presents a recommendation, action-automation is characterized by the human continuously \emph{monitoring and relying} on the automation or intervening to take over control. Thus, the system needs to monitor a user's workload while increasing transparency so that his/her decision is not delayed by overwhelmed workload. 

In this paper, we present 1) an interaction model of human trust-workload dynamics and 2) optimization of the system transparency level based on the estimated human state in a hands-off SAE (Society of Automotive Engineers) level 2 driving context. The driving automation was chosen as it is a promising application of action-automation-based systems.

Level 2 driving automation manages longitudinal and lateral control but requires the driver to monitor the system. We specifically consider a city driving scenario involving multiple intersections, where humans can ``take over'' control, if desired.
One challenge in this application is that unlike the aircraft autopilot system that assumes a trained professional operator, driving automation is expected to be the first safety critical action-automation system that are used by novice users. In this setting, we believe trust calibration is especially important because of two  reasons: 1) the system reliability (at least user's perceived system reliability) is affected by the scene complexity 2) user trust to the system might change frequently depending on system reliability and traffic condition given that most users are not trained experts.   
We develop a probabilistic model of the user trust and workload dynamics using human subject data collected using a driving simulator for urban driving scenes. 
We then optimize system behavior of dynamically varying automation transparency to achieve a better trust-workload tradeoff considering the automation performance.

The contributions can be summarized as follows:
\begin{enumerate}
    \item probabilistic dynamic modeling of human trust-workload behavior in action-automation contexts,
    \item explicit modeling of the coupling between human trust and workload,
    \item driver behavior analysis using time-domain analysis techniques focusing on the effect of transparency on the coupled trust and workload dynamics, and
    \item to the best of our knowledge, this is the first work that optimizes system behavior policy design to calibrate trust in real time for level 2 driving automation considering automation reliability, automation transparency, and scene complexity, along with human reliance and surveillance behavior. 
\end{enumerate}

This paper is organized as follows. Section \ref{sec_background} summarizes related work. Section~\ref{sec_model} describes the proposed framework to \emph{quantitatively} model the dynamics of human trust and workload. The driving simulation study used to collect human subject data and the parameter estimation algorithm are presented in Section~\ref{sec_estimation}. Results and discussion about the estimated model and the corresponding policy are presented in Section~\ref{sec_results}, followed by concluding statements in Section~\ref{sec_conclusion}.

\section{Related Work} \label{sec_background}

Several researchers have developed a variety of human trust models. A large number of these models are qualitative models \cite{moray1999laboratory,muir1994trust,jian2000foundations, desai2012modeling} which analyze the factors that affect trust but cannot be used to make quantitative  predictions. Some quantitative models, including regression models \cite{devries2003effects, muir1996trust} and time-series models of trust \cite{lee1992trust,lee1994trust,moray2000adaptive,lee2004trust,akash2017dynamic,hu2018computational,hoogendoorn2013modelling}, fill this gap but do not account for the probabilistic nature of human behavior. 

Markov models, particularly hidden Markov Models (HMMs) \cite{moe2008learning,malik2009web,elsalamouny2009hmmbased} have been used for probabilistic modeling of trust. While HMMs can incorporate the uncertainty in human behavior \cite{li2003recognition,pineau2003pointbased,wang2009hmm,liu2012modeling}, they do not include the effects of actions from autonomous systems that affect human behavior. An extension of HMMs, partially observable Markov decision processes (POMDPs), provide a framework that does account for these actions and enables the design and synthesis of a policy to choose optimal actions based on a desired reward function. 
POMDPs have been used in HMI for automatically generating robot explanations to improve performance \cite{wang2016trust} and estimating trust in agent-agent interactions \cite{seymour2009trustbased}. Recent work has demonstrated the use of a POMDP model with human trust dynamics to improve human-robot performance~\cite{chen2018planning, akash2019improving,akash2019improving-1,akash2020human}. However, these models do not capture the dynamic effect of automation transparency on human trust-workload behavior in an action-automation context,  specifically in driving automation. In this work, we model human trust-workload behavior as a POMDP and optimally vary automation transparency to improve the interaction between the human and driving automation.

Self-reported surveys are a common tool for assessing a human's trust and workload. Trust surveys include specific questions related to the corresponding experimental context, and a Likert scale is typically used for participants to report their trust in the system~\cite{jian2000foundations}. The NASA TLX survey is the preferred tool to assess human workload~ \cite{proctor2018human}.
In the context of developing algorithms to calibrate human trust in real time, however, it is not practical to use surveys for trust and workload measurements because continuously inquiring humans is not feasible in most contexts. Alternatively, we propose to use behavioral metrics that are readily available in real time and correlate to human trust-workload behavior. For this work, we use human reliance and surveillance behavior through eye-gaze to infer human trust and workload, respectively. The correlation between trust and reliance is well established \cite{parasuraman2008humans,dixon2006automation,lee2004trust}. Furthermore, increased surveillance will lead to increased cognitive load on the human. 
\section{Human Trust-Workload Model} \label{sec_model}

\begin{figure}
\centering
\includegraphics[width=0.47\textwidth]{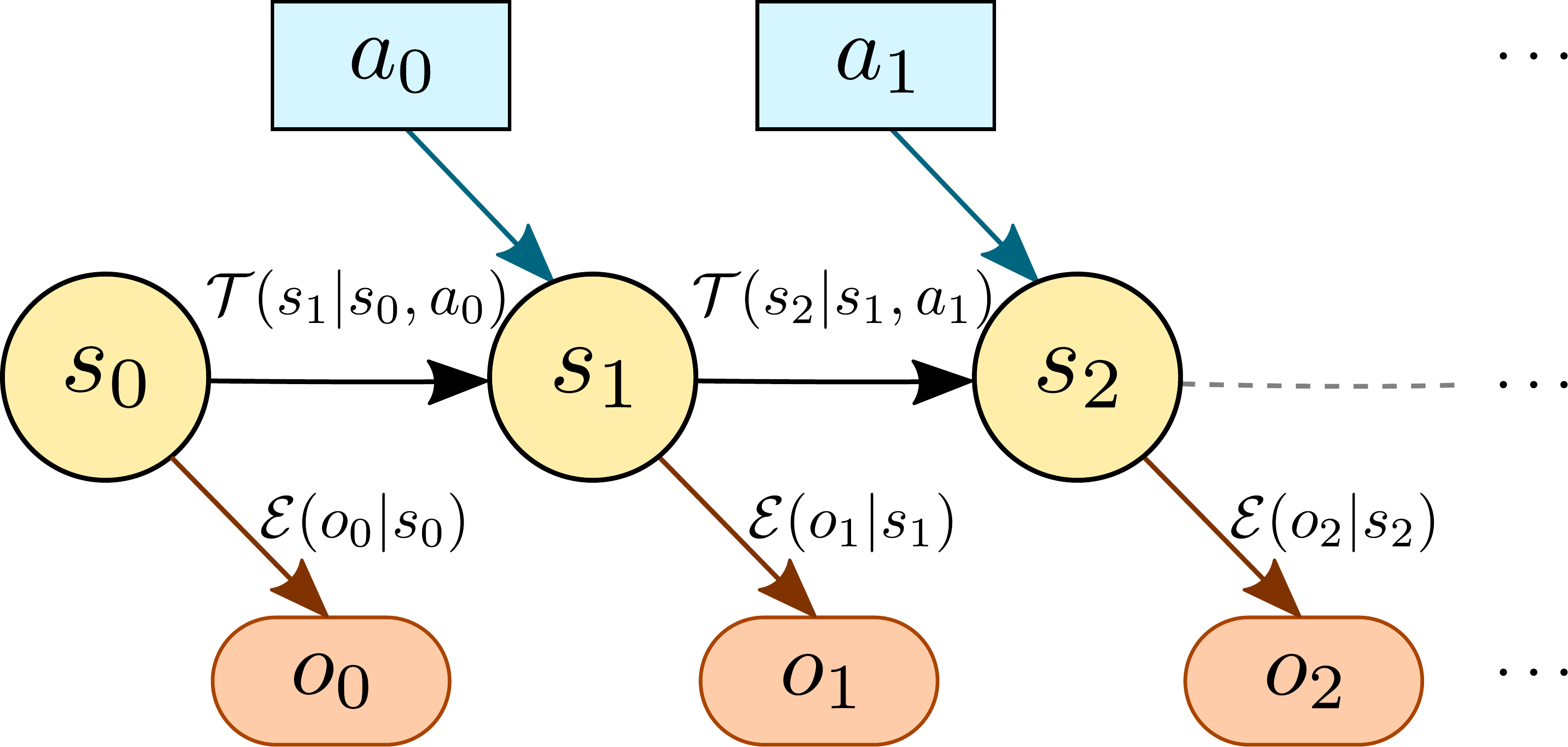}
\caption{A simplified representation of a partially observable Markov decision process (POMDP) model.} 
\label{fig_pomdpstruct}
\end{figure}

Here we propose a probabilistic model for estimating human trust and workload dynamics. We assume that these dynamics follow the Markov property~\cite{puterman2014markov}, and therefore, we model human trust-workload behavior as a POMDP. 
A POMDP is a 7-tuple ($\mathcal{S}, \allowbreak\mathcal{A}, \allowbreak\mathcal{O}, \allowbreak\mathcal{T},  \allowbreak\mathcal{E}, \allowbreak\mathcal{R}, \allowbreak\gamma $) and can be represented as shown in Figure~\ref{fig_pomdpstruct}. Here, $\mathcal{S}$ is a finite set of states, $\mathcal{A}$ is a finite set of actions, and $\mathcal{O}$ is a set of observations.  The transition from the current state $s\in\mathcal{S}$ to the next state $s'\in\mathcal{S}$ given the action $a\in\mathcal{A}$ is characterized by the transition probability function $\mathcal{T}(s'|s,a)$. The emission probability function $\mathcal{E}(o|s)$ characterizes the likelihood of observing $o\in\mathcal{O}$ given the process is in state $s$. Finally, the optimal policy is calculated based on the reward function $\mathcal{R}(s',s,a)$ and the discount factor $\gamma$. Refer to  \cite{sigaud2013markov} for a detailed description of POMDPs.
Many studies have applied POMDPs to model human-machine interaction \cite{akash2020human,akash2019improving,chen2018planning}. A typical allocation is as follows.  $\mathcal{S}$ is associated with the human's internal (i.e., mental) states. Since $\mathcal{S}$ cannot be directly observed by the system, the system action selection is conducted based on a system belief over $\mathcal{S}$, which is estimated through observation $o\in\mathcal{O}$. Based on the belief, system action strategy $\pi$ is optimized to maximize the discounted cumulative reward, where the reward function $\mathcal{R}(s',s,a)$ is designed based on the optimization target of the interaction.

\begin{figure}
\centering
\includegraphics[width=0.47\textwidth]{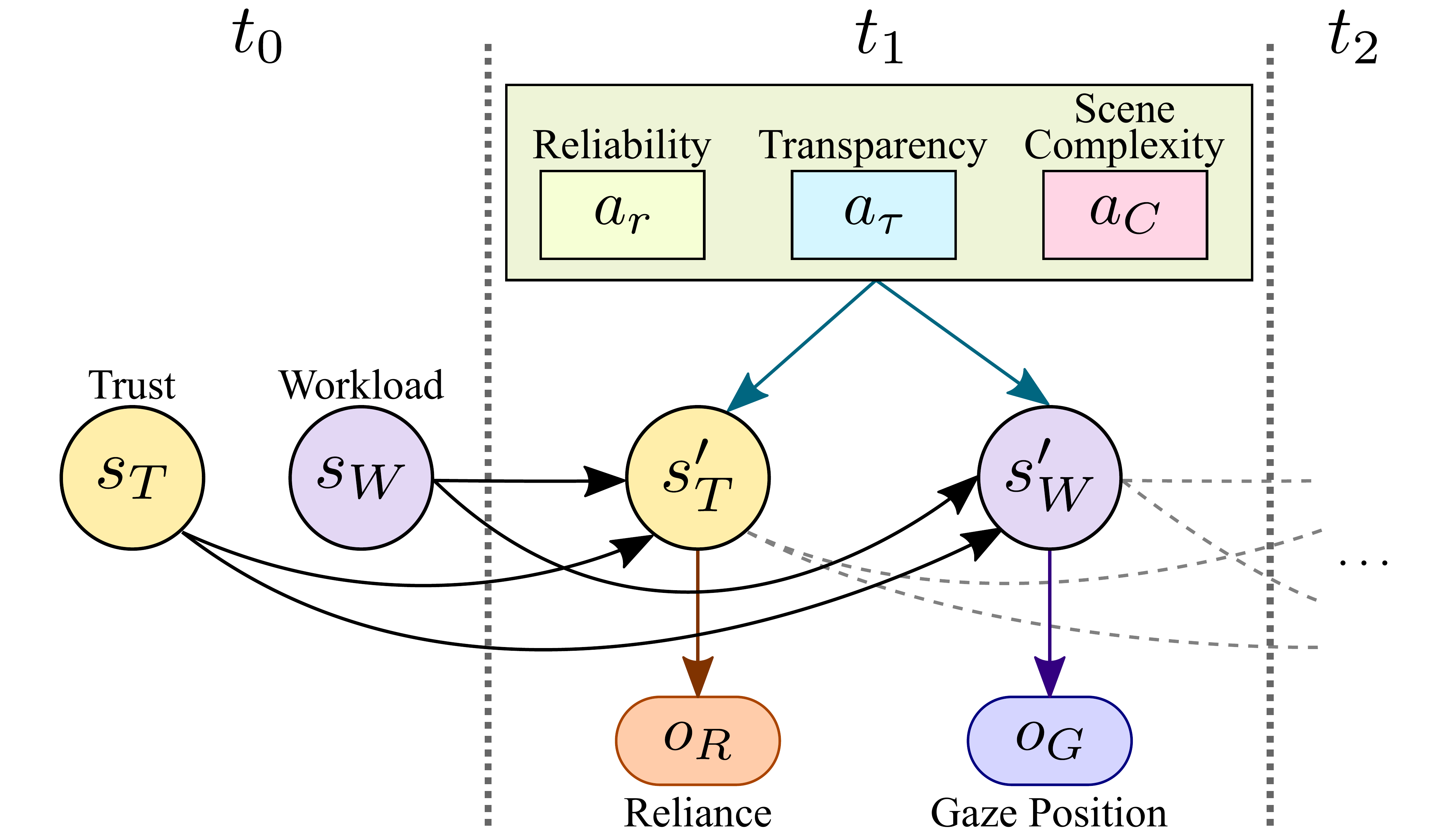}
\caption{The structure of the proposed coupled trust-workload model.}
\label{fig_modelstruct_complete}
\end{figure}

We apply this model to an interaction between a driver and a SAE level 2 driving automation in a ``hands-off'' city-driving scenario. While the level 2 automation is active, the steering, acceleration, and braking are automated. Nevertheless, the human driver has to supervise the automation and ``take over'' control in order to maintain safety as, and when, needed. 
Interaction with level 2 driving automation is characterized by the human's reliance, or lack thereof, on the automation. Furthermore, there is an associated eye-gaze behavior corresponding to the human's supervision of the automation in the environment. We assume that these characteristics of the human's behavior---i.e., reliance and gaze---are dependent on human trust and workload. In particular, we assume that trust only affects reliance, and workload only affects gaze position. This enables the trust and workload states to be identified based on the emission probabilities, which will be discussed later. It should be noted that this assumption is challenged by earlier research that has shown that the relationship between trust and reliance decreases under higher workload \cite{hoff2015trust}. On the other hand, although recent work has shown that there exist correlations between human trust and their gaze behavior \cite{hergeth2016keep, lu2019eye}, it is suspected that intuitive processes that are not captured by self-report measures might have mediated the relationship \cite{hergeth2016keep}.  Given the need to strike a balance between model fidelity and complexity, we assume that any coupled interactions between these particular states and observations can be captured through the coupled interaction between trust and workload; doing so facilitates parameterization of the model as described later in this section. Finally, we assume that human trust and workload are influenced by the characteristics of the automation---reliability and transparency---as well as that of the environment---i.e. scene complexity. The model structure based on these assumptions is illustrated in Figure \ref{fig_modelstruct_complete}.
Table~\ref{tab_model} summarizes the definitions of the model variables.

\begin{table*}
    \caption{Definition of the trust-workload POMDP model. Human trust and workload are modeled as hidden states that are affected by actions corresponding to the characteristics of the automation and the scene. The observable characteristics of the human are modeled as the observations of the POMDP.} \label{tab_model}
    \centering
{\tabulinesep=0.62mm
\begin{tabu}{|l|l|l|}
\hline
\multirow{2}{*}{
$\begin{matrix*}[l] 
\text{States} \\ s 
\in \mathcal{S}
\end{matrix*}$}       & \multirow{2}{*}{
            $s = \begin{bmatrix*}[l] 
            \text{\textit{Trust} } s_T\\ 
            \text{\textit{Workload} } s_W
            \end{bmatrix*}$} & 
                                    $s_T \in    \begin{Bmatrix*}[l]  
                                    \text{Low trust } T_\downarrow, 
                                    \text{High trust } T_\uparrow 
                                    \end{Bmatrix*}$ \\ \cline{3-3} 
& &                                 $s_W \in   \begin{Bmatrix*}[l]    
                                    \text{Low workload } W_\downarrow, 
                                    \text{High workload } W_\uparrow 
                                    \end{Bmatrix*}$ \\ \hline
\multirow{4}{*}{
$\begin{matrix*}[l] 
\text{Actions} \\ a 
\in \mathcal{A}
\end{matrix*}$}       & \multirow{4}{*}{$
            a = \begin{bmatrix*}[l] 
            \text{\textit{Transparency} } a_{\tau} \\ 
            \text{\textit{Reliability} } a_r \\ 
            \text{\textit{Scene Complexity} } a_C 
            \end{bmatrix*}$} & 
                                    $a_{\tau} \in    \begin{Bmatrix*}[l]  
                                    \text{Augmented reality cues absent } \text{AR}_{\text{off}}, 
                                    \text{Augmented reality cues present } \text{AR}_{\text{on}} 
                                    \end{Bmatrix*}$ \\ \cline{3-3} 
& &                                 $a_r \in   \begin{Bmatrix*}[l]  
                                    \text{Low reliability }\text{Rel}_\text{low}, 
                                    \text{Medium reliability }\text{Rel}_\text{mid}, 
                                    \text{High reliability }\text{Rel}_\text{high}
                                    \end{Bmatrix*}$ \\ \cline{3-3} 
& &                                 $a_C \in \text{Traffic density} \times \text{Intersection complexity}$    \\
& &                                 $\text{Traffic density} \coloneqq  \begin{Bmatrix*}[l]  
                                    \text{Low traffic density }\text{Traffic}_\text{low}, 
                                    \text{High traffic density }\text{Traffic}_\text{high}
                                    \end{Bmatrix*}$ \\ 
& &                                 $\text{Intersection complexity} \coloneqq  \begin{Bmatrix*}[l]  
                                    \text{Intersection with only cars }\text{Peds}_\text{absent}, \\
                                    \text{Intersection with both cars and pedestrians }\text{Peds}_\text{present}
                                    \end{Bmatrix*}$ \\ \hline
\multirow{2}{*}{
$\begin{matrix*}[l] 
\text{Observations} \\ o 
\in \mathcal{O}
\end{matrix*}$}       & \multirow{2}{*}{$
            o = \begin{bmatrix*}[l]
            \text{\textit{Reliance} } o_R \\
            \text{\textit{Gaze position} } o_{G} 
            \end{bmatrix*}$} & 
                                    $o_R \in   \begin{Bmatrix*}[l] 
                                    \text{Not relying on automation }R^-, 
                                    \text{Relying on automation }R^+ 
                                    \end{Bmatrix*}$ \\ \cline{3-3} 
& &                                $o_G \in    \begin{Bmatrix*}[l] 
                                    \text{Road }G_\text{road}, 
                                    \text{Vehicle }G_\text{vehi}, 
                                    \text{Pedestrian }G_\text{ped}, 
                                    \text{Sidewalk }G_\text{side}, 
                                    \text{Others }G_\text{oth}
                                    \end{Bmatrix*}$ \\ \hline
\end{tabu}}
\end{table*}

As human trust and workload cannot be directly observed, we define the finite set of states $\mathcal{S}$ of the trust-workload POMDP consisting of tuples of the \textit{Trust} state $s_T$ and the \textit{Workload} state $s_W$, i.e., $s\in\mathcal{S}$ and $s=[s_T,s_W]^T$. The Trust state $s_T$ can either be Low Trust $T_\downarrow$ or High Trust $T_\uparrow$. Similarly, the Workload state $s_W$ can either be Low Workload $W_\downarrow$ or High Workload $W_\uparrow$. Since these hidden states of trust and workload are influenced by the characteristics of the automation and the environment, we define the finite set of actions $\mathcal{A}$ consisting of the tuples $a\in\mathcal{A}$ containing the automation's \textit{transparency} $a_\tau$ and \textit{reliability} $a_r$, along with \textit{scene complexity} $a_C$. The explicit definition of the possible values for each of the actions depends on the specific interaction context and is therefore defined in Section~\ref{sec_estimation} based on the human subject study design considered in this manuscript. The observable characteristics of the human are defined as the finite set of observations $\mathcal{O}$ consisting of human \textit{reliance} $o_R$ and \textit{gaze position} $o_G$. Here, \textit{reliance} $o_R$ can either be the human driver relying on the automation, $o_R=R^+$, or the human driver not relying on the automation and taking over control, $o_R=R^-$. We classify the human driver's \textit{gaze position} $o_G$ at any time belonging to one of five possible values: 1) Road $G_\text{road}$, 2) Vehicle $G_\text{vehi}$, 3) Pedestrian $G_\text{ped}$, 4) Sidewalk $G_\text{side}$, and 5) Others $G_\text{oth}$. Others consists of all other elements in the scene not covered in 1-4, such as the interior of the car, the sky, and buildings.

We assume that at any given time, human trust $s_T'$ and workload $s_W'$ are conditionally independent given the previous states $s_T,s_W$ and actions $a$, i.e., $$\pr(s_T',s_W'|s_T,s_W,a)=\pr(s_T'|s_T,s_W,a)\pr(s_W'|s_T,s_W,a) ,$$ but that trust $s_T$ and workload $s_W$ at the current time affect the next trust state $s_T'$ as well as the next workload state $s_W'$. In this way, the model captures the dynamic coupling between trust and workload behavior as it evolves over time.
This assumption significantly reduces the number of model parameters and in turn, the amount of data needed to estimate them. They also result in separate transition probability functions for trust behavior, $\mathcal{T}_T(s_T'|s_T,s_W,a)$, and workload behavior, $\mathcal{T}_W(s_W'|s_T,s_W,a)$, as well as independent emission probability functions for reliance, $\mathcal{E}_T(o_R|s_T)$, and gaze position, $\mathcal{E}_W(o_G|s_W)$.
\section{Model Parameter Estimation} \label{sec_estimation}

To parameterize the human trust-workload model, we collected human subject data in an experiment designed to analyze the impact of the driving scene on the human driver's {trust in automation}. {This experiment} consisted of a series of interactions with intersections of varying scene complexity accompanied with or without augmented reality (AR) graphical cues. 

\subsection{Human Subject Study}

\emph{Stimuli and Procedure: }A within-subject study was conducted such that each participant completed driving tasks in each of eight ($2\times2\times2$) intersection conditions: two levels of traffic density (low traffic, $\text{Traffic}_\text{low}$, or high traffic, $\text{Traffic}_\text{high}$), two levels of intersection complexity (presence of cars and pedestrians, $\text{Peds}_\text{present}$ or presence of cars only,  $\text{Peds}_\text{absent}$), and two levels of AR cues (annotated, $\text{AR}_\text{on}$, or unannotated, $\text{AR}_\text{off}$). 
The order of the conditions were counterbalanced per participant to reduce expectancy and learning effects over time. The eight driving conditions were organized along two possible routes, with each drive covering 15 blocks consisting of approximately equal left and right turns to ensure a long enough drive without any maneuver-specific responses. In each drive, three intersections were randomly chosen during the experiment design stage that consisted of different numbers of cars and pedestrians based on the drive condition.

High traffic density intersections without pedestrians ($\text{Traffic}_\text{high}+\text{Peds}_\text{absent}$) consisted of eight cars in total with three from each side of the cross traffic and two oncoming cars. Low traffic density intersections without pedestrians ($\text{Traffic}_\text{low}+\text{Peds}_\text{absent}$) consisted of four cars in total with one from each side of the cross traffic and two oncoming cars. High traffic density intersections with the presence of both cars and pedestrians ($\text{Traffic}_\text{high}+\text{Peds}_\text{present}$) consisted of four cars in total with two from each side of the cross traffic (no oncoming) and eight pedestrians crossing the road. Low traffic density intersections with the presence of both cars and pedestrians ($\text{Traffic}_\text{low}+\text{Peds}_\text{present}$) consisted of two cars in total with one from each side of the cross traffic (no oncoming) and four pedestrians crossing the road. 
The AR graphical cues, if present, consisted of bounding boxes surrounding each of the visible cars and pedestrians in the scene. This graphical highlighting was chosen based on the first stage of Endsley's three stage model of situational awareness (i.e. perception) \cite{endsley1995theory}. Cues highlighting pedestrians were marked in blue, while cues highlighting vehicles were marked in red (Figure~\ref{fig_ar_example}). All AR visuals were conformally registered in space to the geospatial center of each visible car/pedestrian and could move through any part of the virtually projected forward road scene with an appearance of being superimposed onto the projected scene.

\emph{Apparatus and Testbed: }The study was conducted in \anon{Virginia Tech's COGnitive Engineering for Novel Technologies (COGENT) laboratory} in a room equipped with a medium-fidelity driving simulator with a fully instrumented cab and approximately 75 degrees of projected virtual canvas at approximately 3 meters in front of the driver eye line. All simulated environments were rendered using Unreal Engine 4.18 \cite{epicgames2019unreal} and enabled detailed visual effects including shadow rendering, post processing, ambient vegetation, and light scattering in high definition. The AR cues were overlaid into the virtual scene in real time via a software developed in Unity \cite{unitytechnologies2019unity}

\emph{Participants: }Sixteen participants (twelve males and four females) from \anon{Virginia Polytechnic Institute and State University} completed the study, ranging in age from 18 to 30 years old. Each participant was required to have a valid driving license for at least two years and to have driven more than 5000 miles per year. None of the participants had experiences with AR-based interfaces. Participants provided informed consent and were briefly given a background of the intended research. 

After being equipped with Tobii Pro 2 eye-tracking glasses, participants completed a practice drive within the virtually simulated city environment. Participants were asked to monitor the autonomous driving mode as it navigated through the urban area. The automated driving was simulated by replaying a past researcher's drive via the ``Wizard of Oz'' technique \cite{wang2017marionette}.  Participants could takeover the automation to ensure continued safety by braking and steering to stop and pull over, respectively. Once a participant felt the situation was safe and released the brake, the automated driving resumed. After the participant felt comfortable with the environment, they completed all eight driving sessions. Each driving session lasted about 4-5 minutes depending on a participant's take-over trials. 
Data from six participants was excluded from the analysis because eye-tracking data could not be recorded completely or with sufficient quality; therefore, ten participants' data was used for estimating the trust-workload model.

\subsection{Estimation}

Although individual differences exist between humans, we assume that a common model can capture the dynamics of human trust and workload behavior{ for the general population}. Therefore, we use the aggregated data from all participants to estimate the transition probability function, observation probability function, and the prior probabilities of states for the trust-workload model. For this study, the automation transparency $a_\tau$ is defined in two levels as the absence of AR annotation cues and the presence of AR annotation cues, i.e., $a_\tau\in\left\{\text{AR}_\text{off}, \text{AR}_\text{on}\right\}$. The scene complexity $a_C$ is characterized by both traffic density ($\text{Traffic}_\text{low}$ or $\text{Traffic}_\text{high}$) and intersection complexity ($\text{Peds}_\text{absent}$ or $\text{Peds}_\text{present}$). We define the automation reliability at an intersection in terms of the distance the driving automation stops the car before the stop line. The automation reliability is defined to be low ($\text{Rel}_\text{low}$) if the car stopped less than 5 meters before, or crossed, the stop line. The reliability is defined to be medium ($\text{Rel}_\text{mid}$) if the car stopped between 5 meters and 15 meters before the stop line, and the reliability is defined to be high ($\text{Rel}_\text{high}$) if the car stopped more than 15 meters away from the stop line. Such a reliability definition is similar to driving aggressiveness, which affects the perceived trustworthiness of the automation \cite{houston2003aggressive,fleiter2010how}. 

Reliance $o_R$ is defined based on the human relying on the automation, $o_R=R^+$, or not relying on the automation and taking over, $o_R=R^-$. Each participant's gaze position is classified as belonging to $G_\text{road}$, $G_\text{vehi}$, $G_\text{ped}$, $G_\text{side}$, or $G_\text{oth}$ in each video frame collected at 25 frames per second. This is achieved by first classifying fixations using Tobii's attention filter with default parameters \cite{olsen2012tobii,tobiipro2015tobii} and then manually annotating each fixation. 
Finally, we assign to all frames after the start of one fixation, and before the start of the next fixation, the annotation of the prior fixation. To estimate the model, we use the data collected during each of the three intersections in which the conditions of interest were varied, along with three seconds before and after the corresponding intersection, in each of the eight drives. 
We define a sequence of action-observation data for each participant as the interaction sequence at each intersection. For the ten participants' data, we have $10\times8\times3$ sequences of data to estimate the parameters of the model.

To estimate parameters of the POMDP model using these sequences of data, we pose and solve an optimization problem to maximize the likelihood of observing the sequences of observation for the given sequences of actions. The Baum-Welch algorithm is typically used to address a similar problem for estimating hidden Markov models (HMMs) (see \cite{rabiner1986introduction} for details). However, HMMs lack the notion of actions; therefore, for estimating the parameters of the trust-workload POMDP model, we use a modified version of the Baum-Welch algorithm that accounts for actions along with the state and observation independence assumptions discussed in Section~\ref{sec_model}. 

\subsection{Model Structure Simplification}
\begin{figure}
\centering
\includegraphics[width=0.48\textwidth]{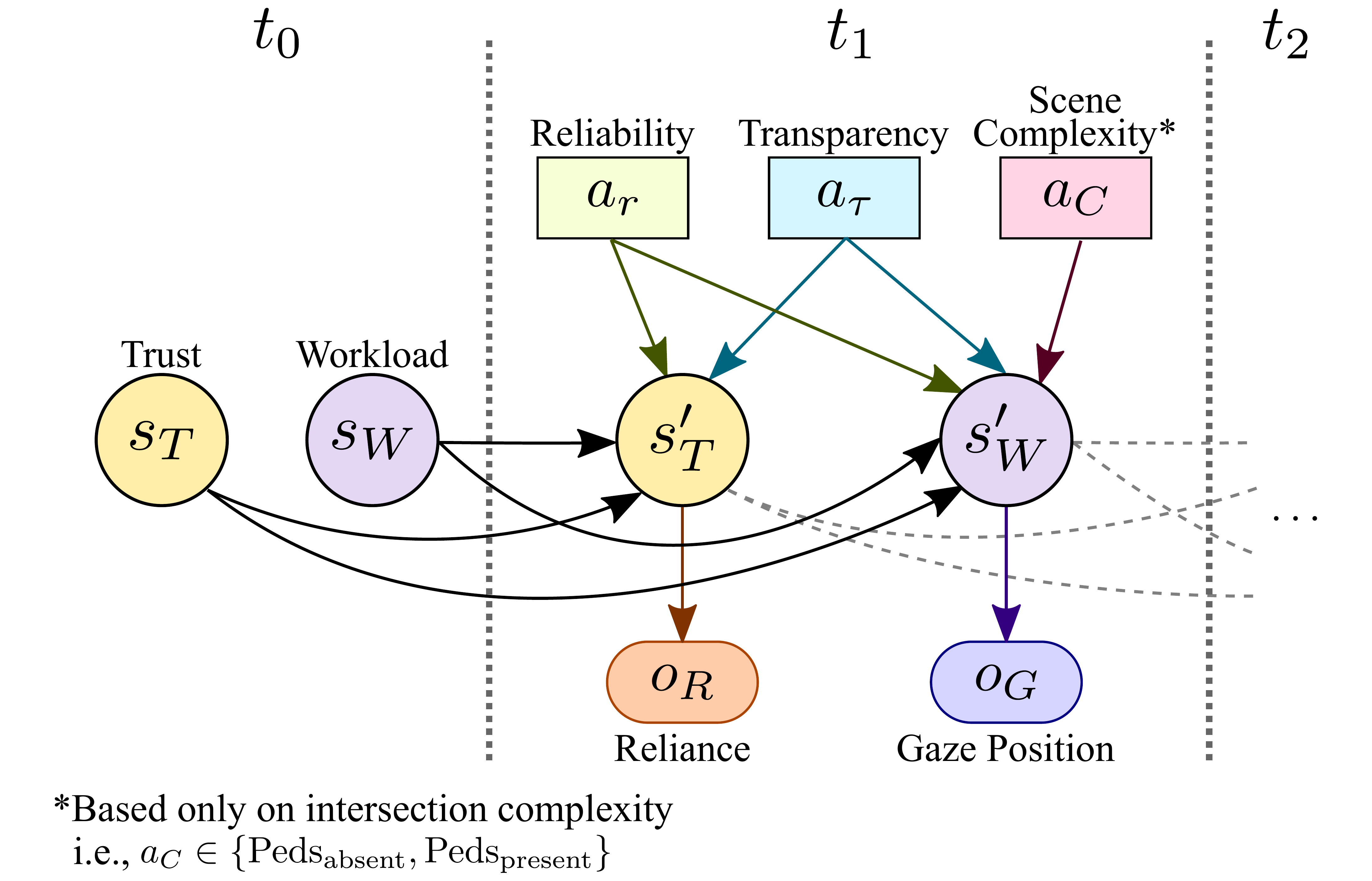}
\caption{The structure of the simplified trust-workload model. The model does not include traffic density as an action and the scene complexity based on intersection complexity affects only the workload state.}
\label{fig_modelstruct_simplified}
\end{figure}

To obtain a model with the best generalizability given the available data, we find the \emph{subset} of actions that directly affect the trust and workload dynamics. For example, we fix reliability to always be an action for the trust dynamics as it has been established that reliability affects trust. 
We then train all possible trust-workload models with different subsets of actions for trust and workload. We conduct a 3-fold cross validation for each possible model, with each model trained 24 times with a different division of training and testing sets to reduce uncertainty in the estimated validation likelihood. We ensure that each fold contains one intersection from each condition of the experiment for each participant to maintain uniformity of the data across the folds. Furthermore, we calculate the average validation likelihood for each of the models and select the model that minimizes the Akaike information criterion (AIC) for the average validation likelihood \cite{burnham2002model}. The resulting model consists of automation reliability and automation transparency as actions for trust dynamics, and automation reliability, automation transparency, and intersection complexity as actions for workload dynamics. The model does not include traffic density as an action, which agrees with the findings based on the questionnaire \cite{wu2020drivers} in which traffic density was found to be insignificant. The resulting model structure is represented in Figure~\ref{fig_modelstruct_simplified} and has significantly less parameters as there would be in a model that had not been simplified as described here. It also has the maximum average validation likelihood among all models; therefore, this model structure generalizes well. 

Finally, the entire dataset is used to estimate the parameters of the model structure represented in Figure~\ref{fig_modelstruct_simplified}. In order to avoid local minima in parameter estimation, we iterate the {algorithm} 1000 times, with each iteration using a different initial guess of the parameters. The estimated POMDP model of trust-workload behavior is presented and analyzed in the next section. 

\section{Results and Discussion} \label{sec_results}

\subsection{Analysis of driver behavior}
The estimated model consists of initial state probabilities for trust $\pi(s_T)$ and workload $\pi(s_W)$, emission probability functions $\mathcal{E}_T(o_R|s_T)$ and $\mathcal{E}_W(o_G|s_W)$, and transition probability functions $\mathcal{T}_T(s_T'|s_T,s_W,a)$ and $\mathcal{T}_W(s_W'|s_T,s_W,a)$. Based on the emission probability function for trust $\mathcal{E}_T(o_R|s_T)$, we define the High Trust state $s_T=T_\uparrow$ as that in which there is a higher probability of observing the human rely on the automation, $o_R=R^+$. Based on the emission probability function for workload $\mathcal{E}_W(o_G|s_W)$, we define the state with the higher entropy of emission probability as the High Workload state $s_W=W_\uparrow$. The entropy $S(s_W)$ for the workload state $s_W$ is calculated as \cite{cover2006elements}
$$ S(s_W) = -\sum\limits_{o_G} \mathcal{E}_W(o_G|s_W) \log\left(\mathcal{E}_W(o_G|s_W)\right) \enspace .$$ 

\begin{figure}
\centering
\includegraphics[width=0.35\textwidth]{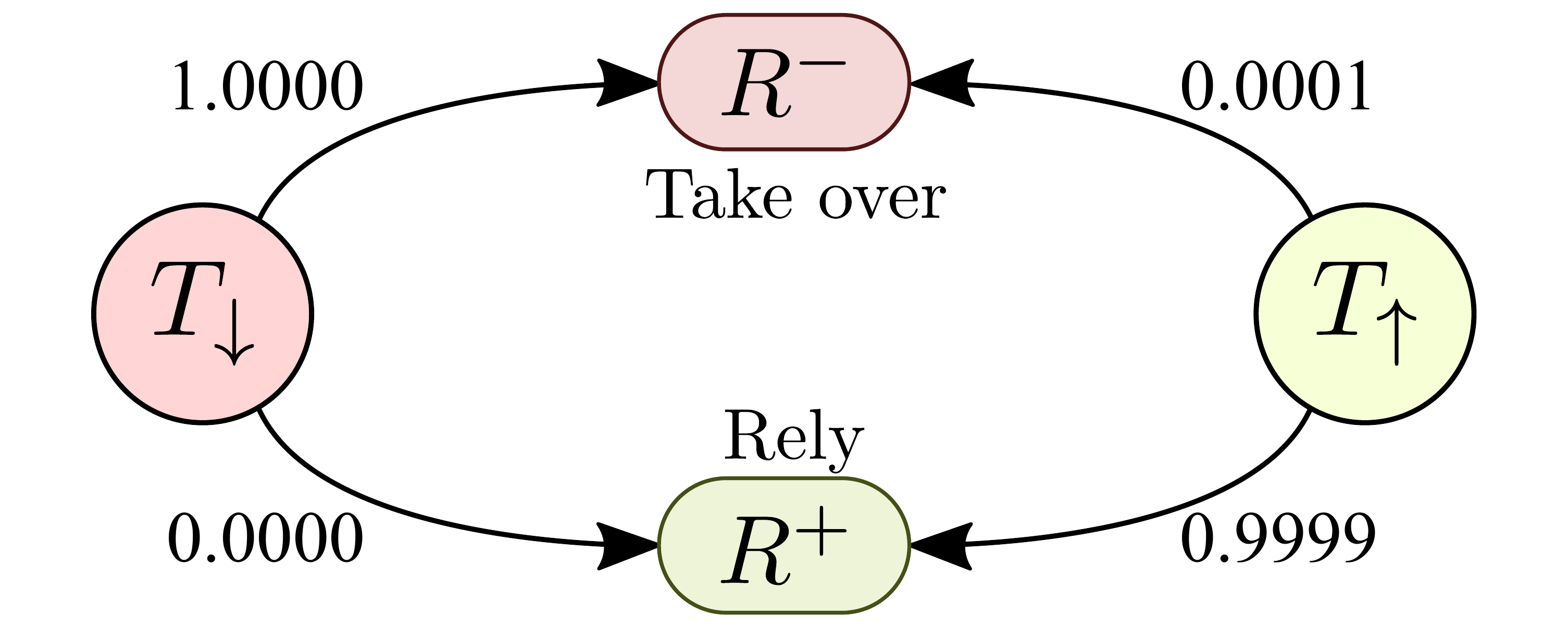}
\caption{Emission probability function $\mathcal{E}_T(o_R|s_T)$ for reliance. Probabilities of observation are shown beside the arrows.}
\label{fig_emission_T}
\end{figure}

The estimated initial probabilities of Low Trust $T_\downarrow$ and High Trust $T_\uparrow$ are $\pi(T_\downarrow) = 0.0000$ and $\pi(T_\uparrow) = 1.0000$, respectively. 
The emission probability function $\mathcal{E}_T(o_R|s_T)$ is depicted in Figure~\ref{fig_emission_T} and characterizes the probability of the participants' reliance on the automation given the participants' state of trust. When in a state of Low Trust, the likelihood of participants not relying on the automation is one.  Similarly, when in a state of High Trust, the likelihood of participants relying on the automation is nearly one.

\begin{figure}
\centering
\includegraphics[width=0.35\textwidth]{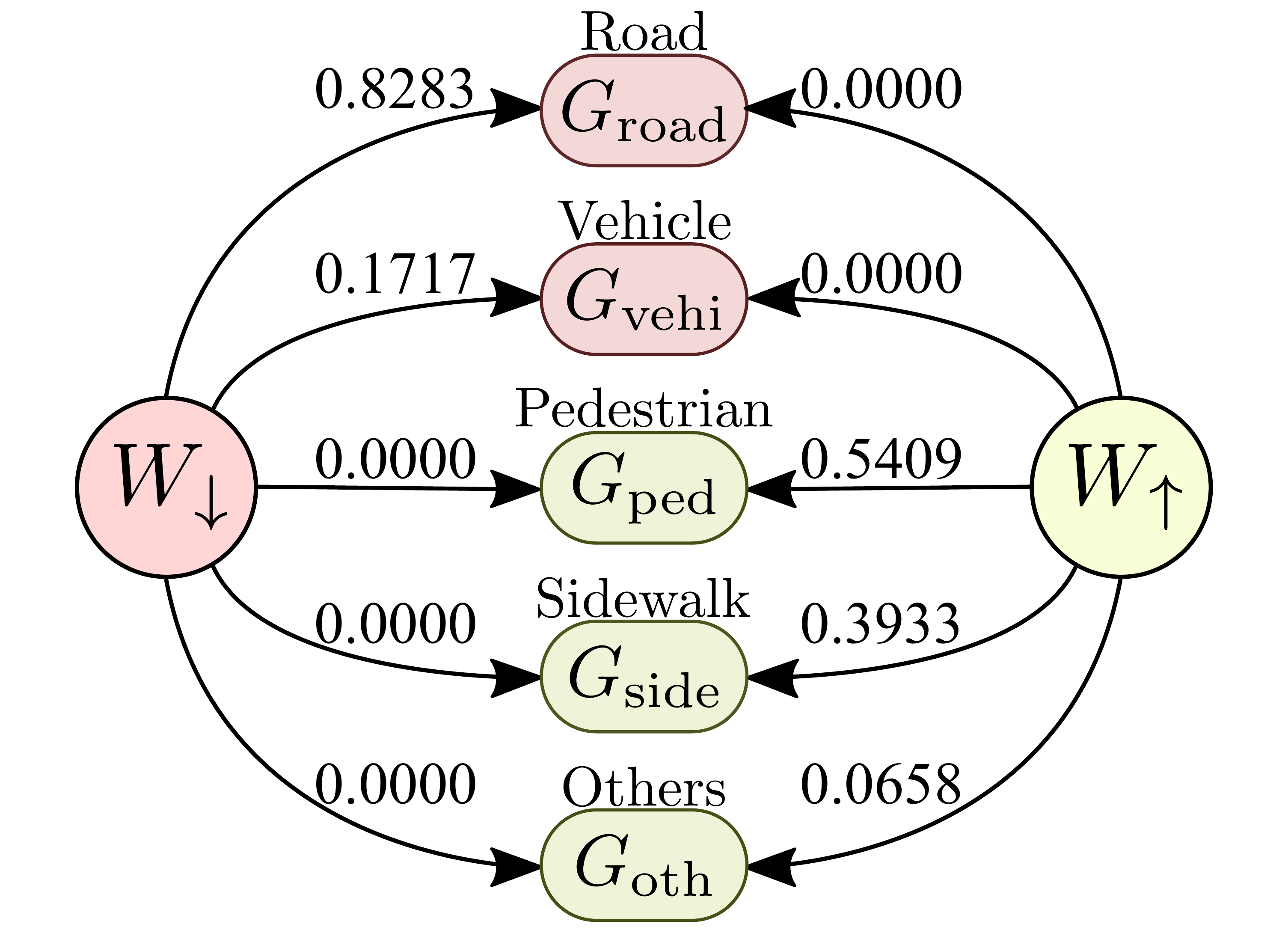}
\caption{Emission probability function $\mathcal{E}_W(o_G|s_W)$ for gaze position. Probabilities of observation are shown above the arrows.}
\label{fig_emission_W}
\end{figure}

The estimated initial probabilities of Low Workload $W_\downarrow$ and High Workload $W_\uparrow$ are $\pi(W_\downarrow) = 0.5833$ and $\pi(W_\uparrow) = 0.4167$, respectively. The emission probability function $\mathcal{E}_W(o_G|s_W)$ is depicted in Figure~\ref{fig_emission_W} and characterizes the probability of the participants' gaze position on the scene given the participants' state of workload. For the Low Workload state, participants focus more on the road and the vehicles on the road. However, for the High Workload state, participants' focus is distributed between pedestrians and the sidewalk, along with other elements in the scene.

\begin{figure*}
\centering
\subfigure[$a_C=\text{Peds}_\text{absent}, a_r=\text{Rel}_\text{low}$.\label{fig_step2_3}]{\includegraphics[width=0.33\textwidth]{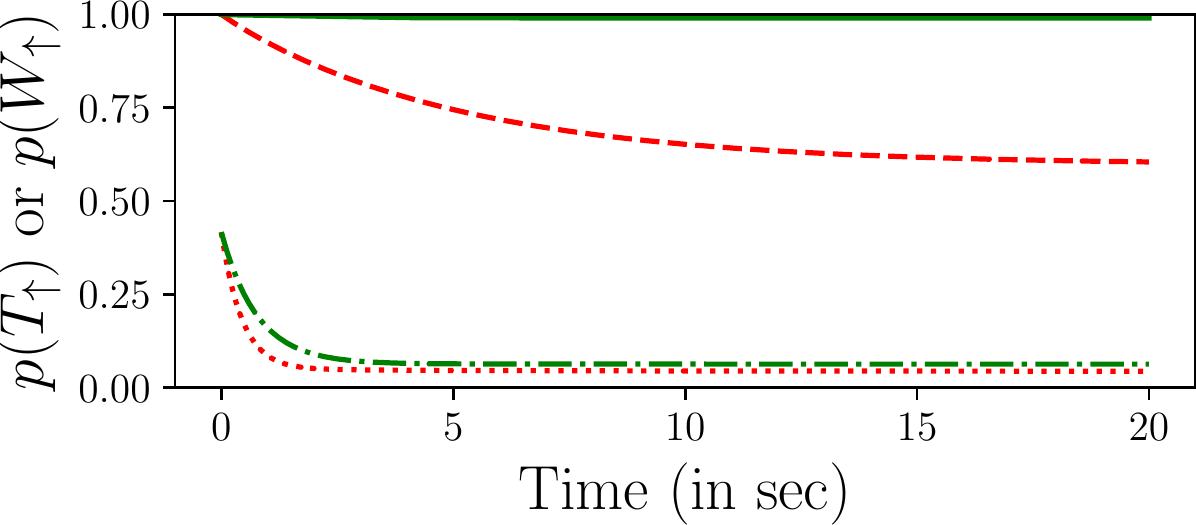}}\hfill
\subfigure[$a_C=\text{Peds}_\text{absent}, a_r=\text{Rel}_\text{mid}$.\label{fig_step4_5}]{\includegraphics[width=0.33\textwidth]{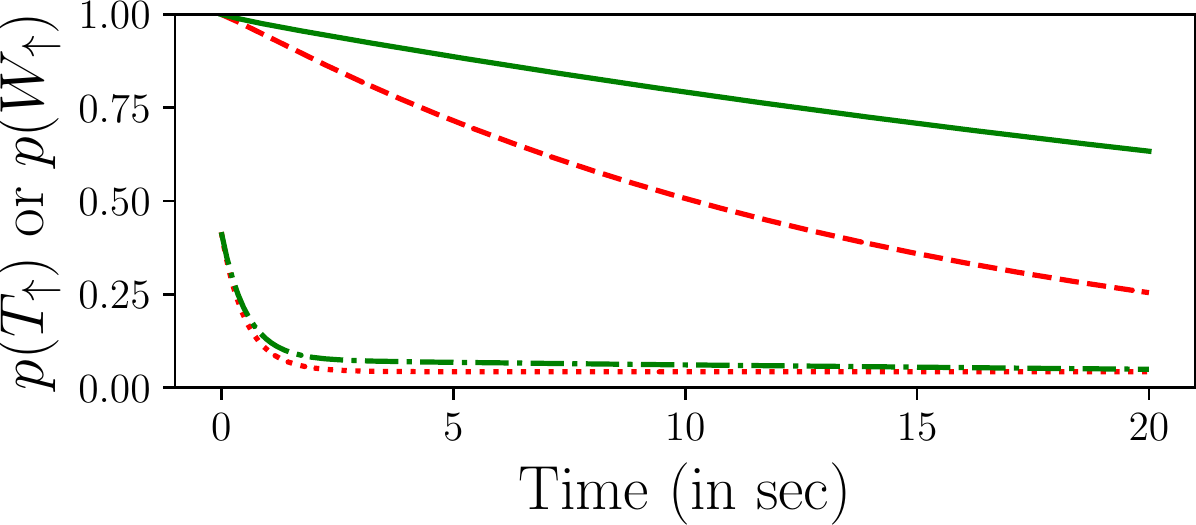}}\hfill
\subfigure[$a_C=\text{Peds}_\text{absent}, a_r=\text{Rel}_\text{high}$.\label{fig_step6_7}]{\includegraphics[width=0.33\textwidth]{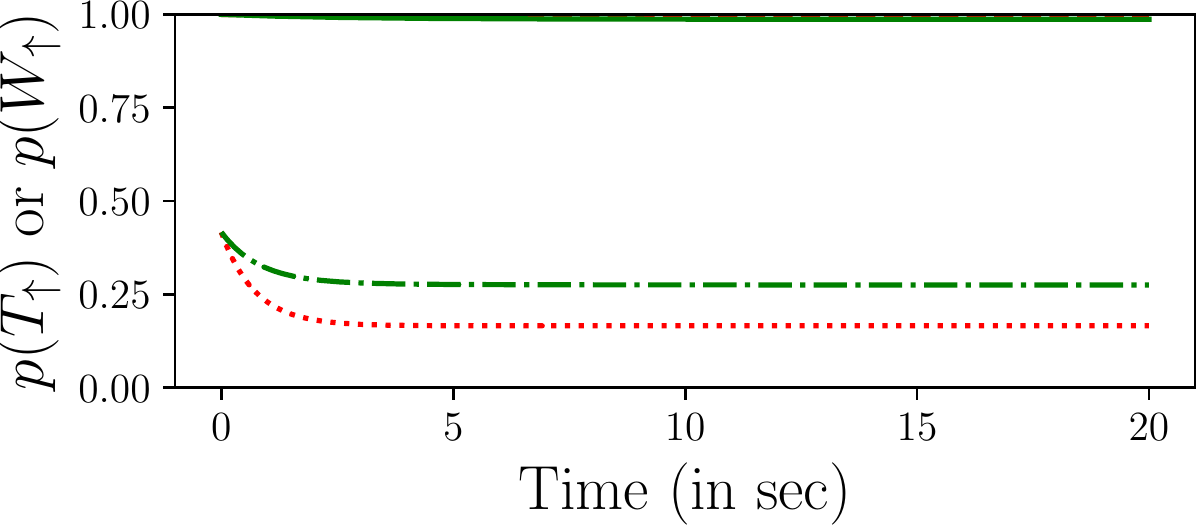}}\\ 
\subfigure[$a_C=\text{Peds}_\text{present}, a_r=\text{Rel}_\text{low}$.\label{fig_step8_9}]{\includegraphics[width=0.33\textwidth]{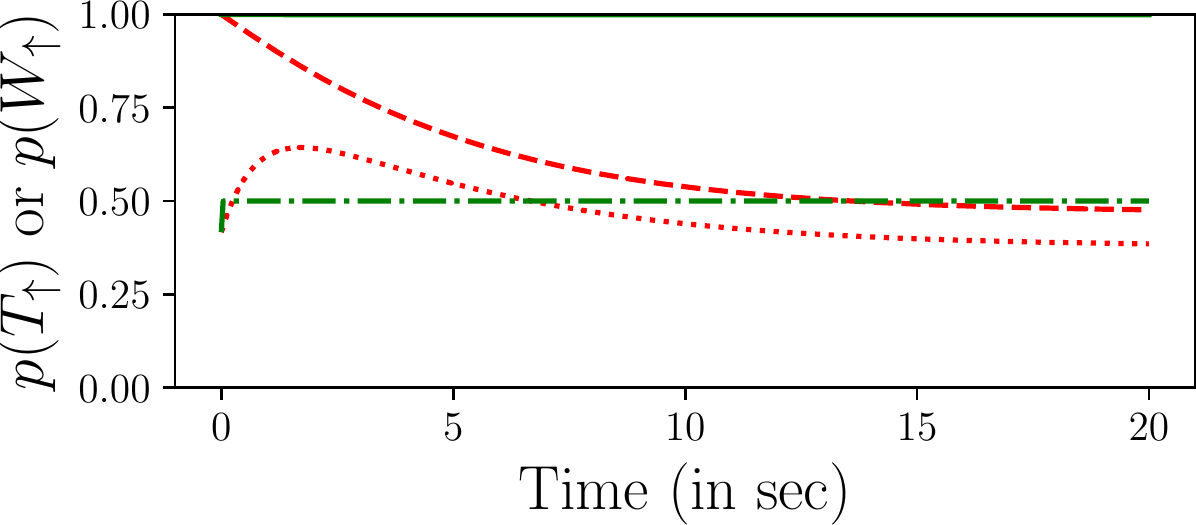}}\hfill
\subfigure[$a_C=\text{Peds}_\text{present}, a_r=\text{Rel}_\text{mid}$.\label{fig_step10_11}]{\includegraphics[width=0.33\textwidth]{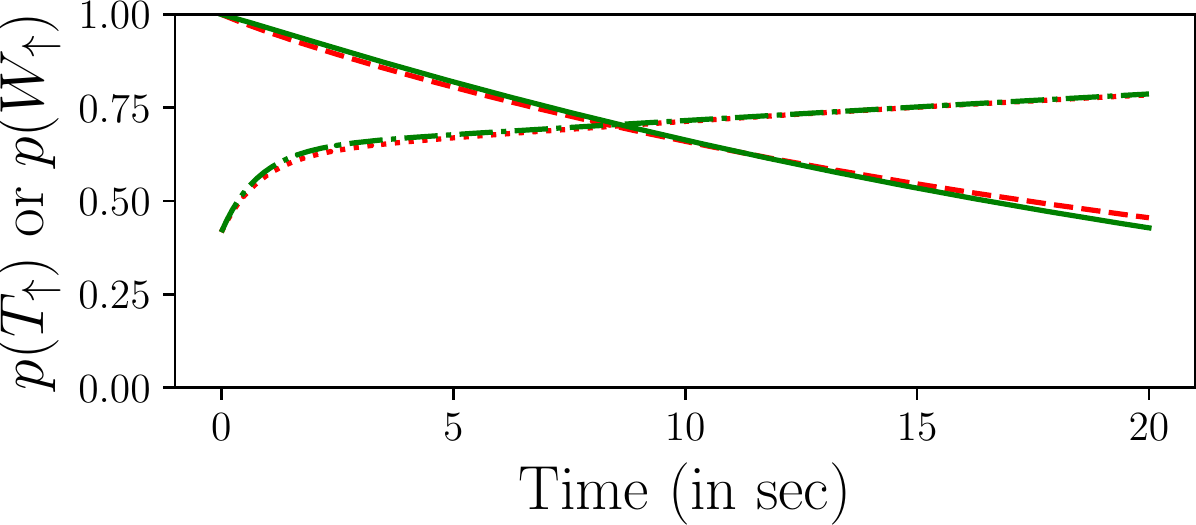}}\hfill
\subfigure[$a_C=\text{Peds}_\text{present}, a_r=\text{Rel}_\text{high}$.\label{fig_step12_13}]{\includegraphics[width=0.33\textwidth]{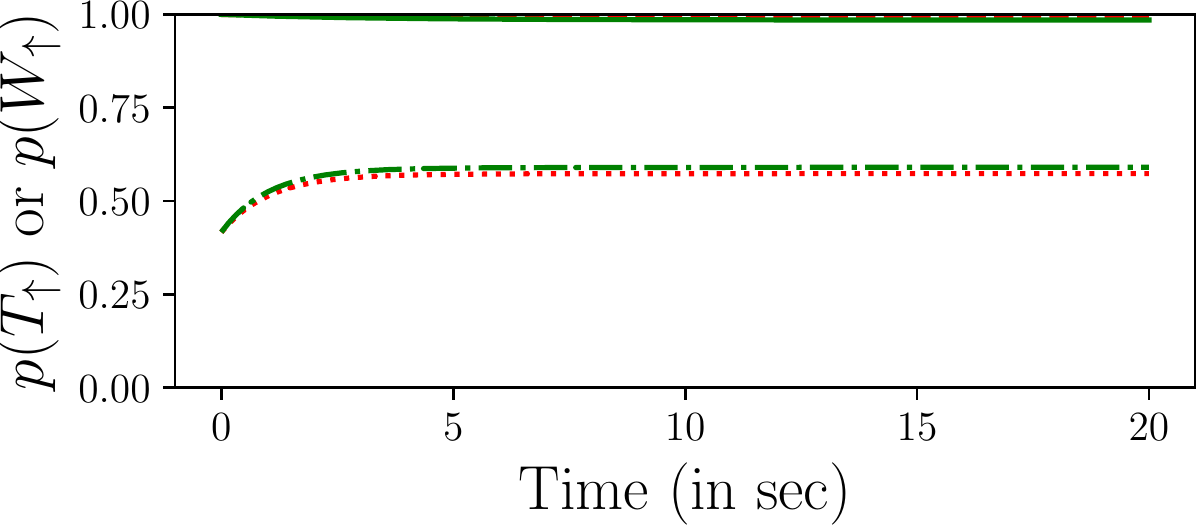}}
\subfigure{\includegraphics[width=0.33\textwidth]{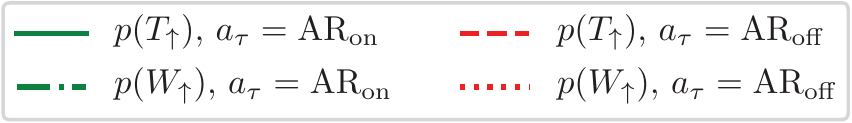}}
\caption{Step responses of human state of trust and workload for each of the actions of the POMDP model.}
\label{fig_stepresponses}
\end{figure*}

To analyze how the actions---transparency, reliability, and scene complexity---affect the state dynamics over time, we simulate step responses as shown in Figure ~\ref{fig_stepresponses}. Here, a step response for an action $a$ can be construed as the evolution of the probability that the human is in a state of High Trust $T_\uparrow$ and High Workload $W_\uparrow$ as the POMDP evolves under the given action. 
Each of the plots in Figure~\ref{fig_stepresponses} compare the step response of trust and workload between the two transparency levels---absence and presence of AR cues---for a given scene complexity, $a_C$, and automation reliability, $a_r$. We observe that over time, in most cases, the probability of High Trust decayed faster to an equal or lower value if AR cues were absent (red dashed lines) as compared to when AR cues were present (green solid lines). This is consistent with the findings based on the questionnaire \cite{wu2020drivers} in which more participants thought the system with the AR cues could provide advice for their decision making compared to the system without AR cues. Also, the probability of High Workload converged to a lower or equal value when AR cues were absent (red dotted lines) as compared to when AR cues were present (green dot-dashed lines). 

Furthermore, Figures~\ref{fig_step6_7} and \ref{fig_step12_13} show that high automation reliability saturates the probability of High Trust to a very high level irrespective of the scene complexity and automation transparency. This is expected given that reliability has been shown to strongly impact human trust. Nonetheless, we observe that high automation transparency (presence of AR cues) is able to maintain high trust even in low reliability cases (Figures~\ref{fig_step2_3} and \ref{fig_step8_9}) because the participants are able to make more informed decisions. For medium reliability cases, shown in (Figures~\ref{fig_step4_5} and \ref{fig_step10_11}), participants' trust decreases over time, possibly because the participants may be unsure of the trustworthiness of the system given that the car neither stops too close to the stop sign nor far enough. Considering the workload state, we observe that high scene complexity (presence of pedestrians at the intersection) results in higher probability of High Workload (Figures~\ref{fig_step8_9}, \ref{fig_step10_11}, and \ref{fig_step12_13}) as compared to low scene complexity (Figures~\ref{fig_step2_3}, \ref{fig_step4_5}, and \ref{fig_step6_7}). 

To summarize, the parameterized model captures the expected trade-off: that increasing transparency can increase trust but also increases cognitive workload. Moreover, the results suggest that the effect of increased transparency on human trust and workload also depends on other factors including, but not limited to, scene complexity and automation reliability.

\subsection{Analysis of optimized policy}
Next, we obtain the optimal policy aimed at calibrating trust.
The obtained trust-workload model provides the ability to estimate trust and workload levels of a human continuously, and in real time, using the belief state estimates \cite{spaan2012partially}.
In order to obtain a policy that can calibrate human trust, we define a reward function {as a function of the human trust state $s_T$ and automation reliability $a_r$ }as shown in Table~\ref{tab_reward}.  We allocate a penalty of $-1.0$ when the model predicts that the human is in a state of High Trust given low automation reliability or when it predicts that the human is in a state of Low Trust given high automation reliability. 
We allocate a reward of $1.0$ when the model predicts the human is in a state of High Trust given high automation reliability and when it predicts the human is in a state of Low Trust given low automation reliability.
\begin{table}
\caption{Reward function used to calibrate human trust.}
\label{tab_reward}
\small
\centering
\begin{tabular}{l r r r} 
 \hline
 \hline
        	& $\text{Rel}_\text{low}$ 	& $\text{Rel}_\text{mid}$ 		& $\text{Rel}_\text{high}$  \\ 
 \hline
 Low Trust $T_\downarrow$   		& $1.0$ 	& $0.0$ 	& $-1.0$  \\
 High Trust $T_\uparrow$   		& $-1.0$ 		& $0.0$ 		& $1.0$ \\
 \hline
 \hline
\end{tabular}
\end{table}
We select the discount factor $\gamma$ such that the reward after one second has a weight of $e^{-1}$; given 25 time steps per second for our dataset, such a value of $\gamma$ can be approximated as $\gamma= \frac{25}{25+1} = 0.9615$.
With this reward function and discount factor, we calculate the optimal policy for the POMDP to determine the optimal system transparency level.

Although the exact optimal solution for a POMDP can be obtained using value iteration, the time complexity of solving a POMDP via value iteration grows exponentially with an increase in the cardinality of the action and observation sets. In real-life scenarios the sets of actions and observations can be much larger; therefore, using exact value iteration is infeasible. Instead, we use the Q-MDP method to obtain a near-optimal solution \cite{cassandra1994acting}. Furthermore, to account for the uncontrollable actions that cannot be explicitly changed by the automation (i.e. reliability and scene complexity), we calculate the expected Q-function by considering the probabilities of the uncontrollable actions as discussed in \cite{akash2019improving}. Here we consider that all uncontrollable actions are equiprobable. In a real scenario, prior route and automation knowledge could be used to determine the probabilities of the uncontrollable actions. 

\begin{figure}
\centering
\subfigure[$\text{Peds}_\text{absent}$ and $\text{Rel}_\text{low}$.\label{fig_CO_relLOW}]{\includegraphics[width=0.22\textwidth]{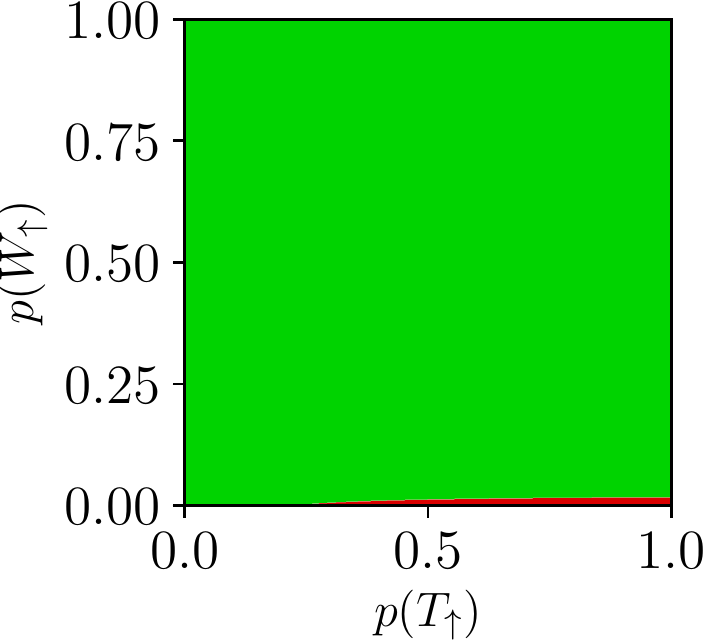}}
\subfigure[$\text{Peds}_\text{present}$ and $\text{Rel}_\text{low}$.\label{fig_CP_relLOW}]{\includegraphics[width=0.22\textwidth]{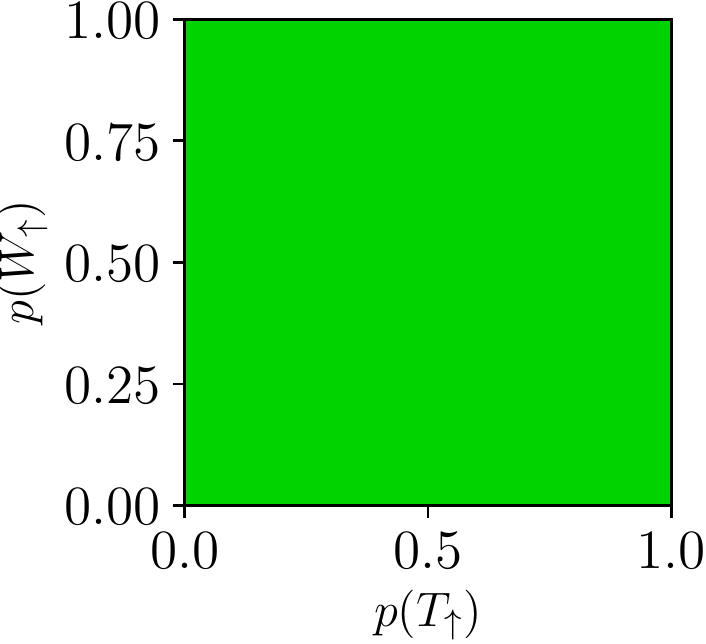}}\\
\subfigure[$\text{Peds}_\text{absent}$ and $\text{Rel}_\text{mid}$.\label{fig_CO_relMID}]{\includegraphics[width=0.22\textwidth]{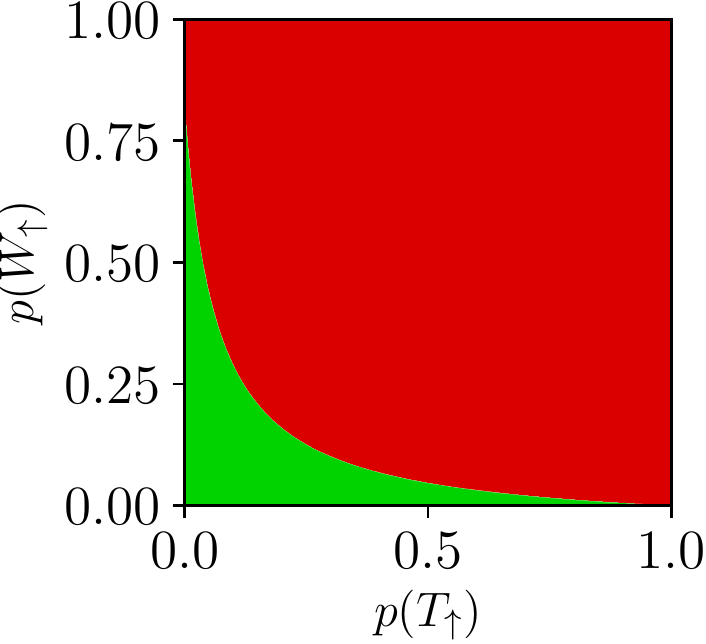}}
\subfigure[$\text{Peds}_\text{present}$ and $\text{Rel}_\text{mid}$.\label{fig_CP_relMID}]{\includegraphics[width=0.22\textwidth]{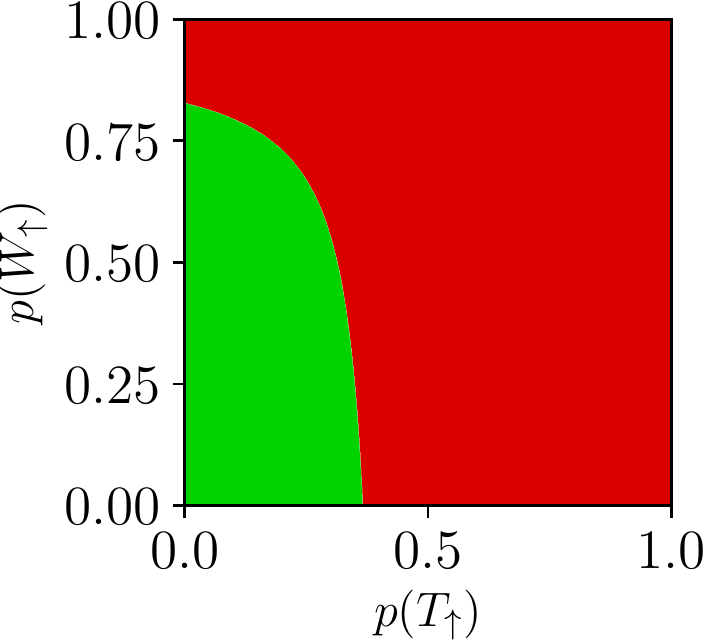}}\\
\subfigure[$\text{Peds}_\text{absent}$ and $\text{Rel}_\text{high}$.\label{fig_CO_relHIGH}]{\includegraphics[width=0.22\textwidth]{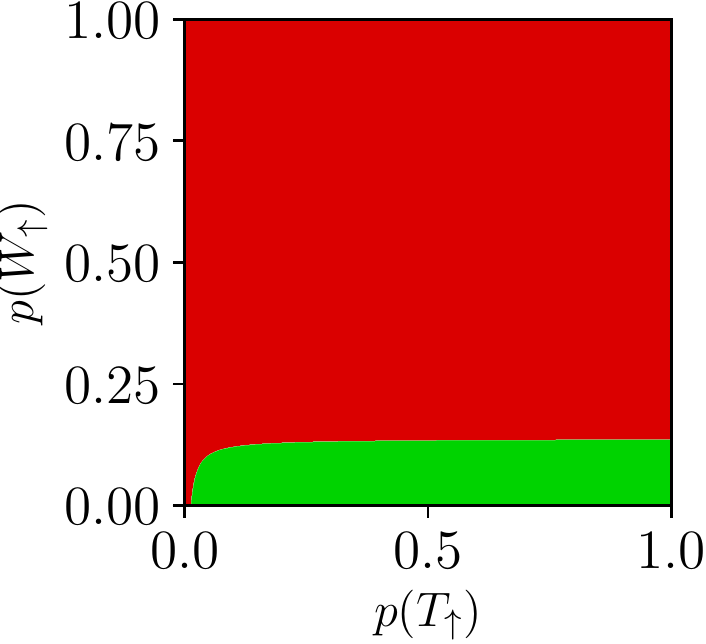}}
\subfigure[$\text{Peds}_\text{present}$ and $\text{Rel}_\text{high}$.\label{fig_CP_relHIGH}]{\includegraphics[width=0.22\textwidth]{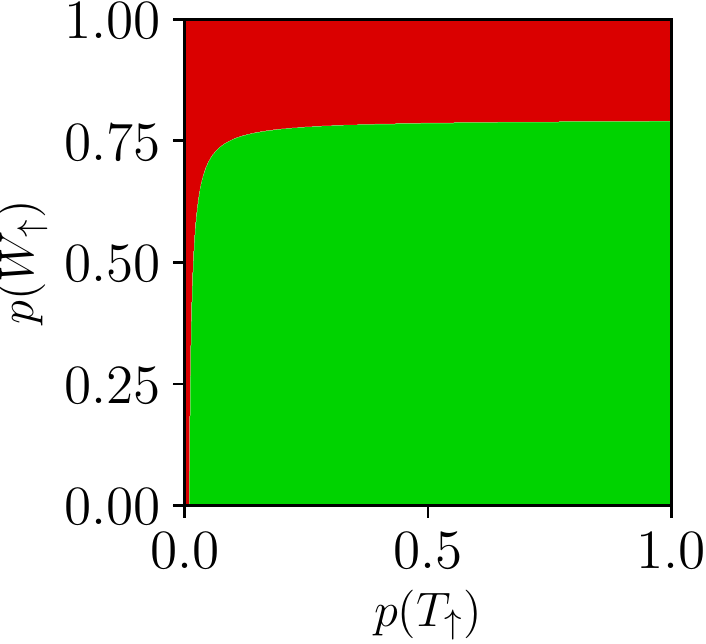}}\\
\subfigure{\includegraphics[width=0.15\textwidth]{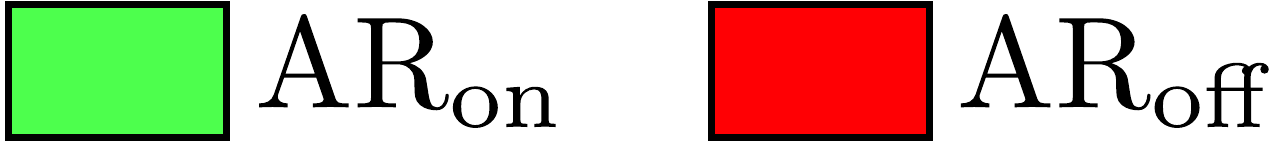}}
\caption{Optimized policy for calibrating human trust. The color indicate if the system should turn-on the AR cues (green) or turn-off the AR cues (red) depending on system's belief on user's trust and workload state. }
\label{fig_control}
\end{figure}

The optimal policy to select the action (transparency $a_\tau$) corresponding to each of the uncontrollable actions (reliability $a_r$ and scene complexity $a_C$) is depicted in Figure~\ref{fig_control}. It is worth noting that even though the reward function is defined in terms of the trust state $s_T$, the system action is also dependent on the workload state $s_W$. For example, the system learned to reduce the driver workload when system reliability is medium or high and the scene is not very complicated by adopting absence of AR cues ($\text{AR}_\text{off}$) (see Figures~\ref{fig_CO_relMID} and \ref{fig_CO_relHIGH}); this is due to the coupled modeling of trust and workload. For low reliability cases (Figures~\ref{fig_CO_relLOW} and \ref{fig_CP_relLOW}), the optimal policy adopts the presence of AR cues ($\text{AR}_\text{on}$) as the transparency level. This high transparency will allow the human to make an informed decision and avoid mistrust. For medium reliability cases (Figures~\ref{fig_CO_relMID} and \ref{fig_CP_relMID}), the optimal policy adopts high transparency ($\text{AR}_\text{on}$) when both trust and workload are low. Providing high automation transparency at the low trust state helps to increase the human's trust, but it is avoided when the human's workload is high. Similarly, for high reliability cases (Figures~\ref{fig_CO_relHIGH} and \ref{fig_CP_relHIGH}), high transparency is only used when the human's workload is low. Interestingly, high transparency ($\text{AR}_\text{on}$) is adopted even when the human's workload is high when pedestrians are present (Figures~\ref{fig_CP_relLOW}, \ref{fig_CP_relMID}, and \ref{fig_CP_relHIGH}). One potential reason for this is that the presence of pedestrians may be interpreted as ``higher risk'' to the human, thereby leading to less trust in the automation if AR cues are absent. However, identification of potential confounding effects of risk is out of the scope of this work. Nonetheless, these results highlight the importance of including factors such as scene complexity, in addition to automation reliability in such a model used for real-time decision making.

In summary, the trust and workload-based feedback policy described here provides a framework for achieving adaptive transparency based on a quantitative dynamic model of human behavior. Although based on a limited sample size of human subject data, the framework provides an insight toward the coupled interactions between human trust and workload and how the corresponding dynamics can be exploited to optimally calibrate trust to improve human-automation interactions.

\section{Conclusion} \label{sec_conclusion}

We presented a POMDP framework to model coupled human trust and workload dynamics as they evolve during a human's interaction with a hands-off SAE level 2 driving automation. The model was trained using human subject data, collected via a medium-fidelity driving simulator with variations in scenarios that captured the effects of automation reliability, automation transparency, and scene complexity, along with reliance and eye-gaze behavior, on the dynamics of human trust and workload. Analysis of the model demonstrates that user behavior is strongly influenced by the scene complexity and it should be accordingly considered when determining the optimal transparency. Using a reward function designed to calibrate trust, we obtained an optimal policy to achieve trust calibration. The proposed algorithm can influence driver trust level and workload by controlling automation transparency level depending on the human's current trust and workload level along with automation reliability and scene complexity. While we trained a ``general'' policy that applies to all participants, which mostly comprised of young adults, the optimal policy might also depend on the individual factors of each driver. We would like to carefully investigate the effect of these driver-dependent factors when we conduct a validation of this policy in real time with human subjects. Finally, it is worth noting that real driving is more complex than the scenarios we created; therefore, the ecological validity is limited. We believe that research to explore the effects of attributions to driver trust behavior and to influence user trust and workload level will provide essential and necessary steps towards developing human-aware automated system interfaces. 
\begin{acks}

We sincerely acknowledge COGNET lab at Virginia Polytechnic Institute and State University for the collection of the human subject data used in this paper. We also thank Yuki Gorospe at Honda Research Institute US, Inc. for helping with eye-gaze annotation.

\end{acks}

\bibliographystyle{ACM-Reference-Format}
\bibliography{Trust_Group_Lib_References,sample-base}










\end{document}